# Ensemble Consider Kalman Filtering*

Tai-shan Lou, Nan-hua Chen, Hua Xiong, Ya-xi Li, and Lei Wang

*Abstract*—For the nonlinear systems, the ensemble Kalman filter can avoid using the Jacobian matrices and reduce the computational complexity. However, the state estimates still suffer greatly negative effects from uncertain parameters of the dynamic and measurement models. To mitigate the negative effects, an ensemble consider Kalman filter (EnCKF) is designed by using the "consider" approach and resampling the ensemble members in each step to incorporate the statistics of the uncertain parameters into the state estimation formulations. The effectiveness of the proposed EnCKF is verified by two numerical simulations.

## I. INTRODUCTION

Nonlinear Kalman filtering plays an important role in information and communication systems, control systems, and many other areas, such as target tracking, navigation of aerospace vehicles[1, 2], fault diagnosis[3], chemical plant control, signal processing and fusion of multi-sensor data[4, 5]. A lot of Kalman filtering algorithms have been proposed to different engineering problems, such as extended Kalman filter, central differential Kalman filter, unscented Kalman filter, cubature Kalman filter and ensemble Kalman filter (EnKF)[6]. The unscented Kalman filter and cubature Kalman filter belong to the deterministic sampling filter algorithm, in which the sigma points are generated deterministically on the state and covariance matrix[6]. The EnKF belongs to a general class of known particle algorithm, in which an ensemble is used to represent for the probability distribution functions (PDFs), the time-update PDFs and the posterior PDF of the measurements are modeled by the stochastic models of the ensemble integration, respectively. The EnKF is widely used in the nonlinear models with the extremely high order, high uncertainty of the initial states and a large number of observations[7].

For the above nonlinear Kalman filtering, an underlying assumption is that the dynamic and measurement equations can be accurately modeled without any unknown parameters or biases. However, in practice, it is always difficulty to obtain the accurately parameter values, and sometimes the parameters are time-varying[8]. Neglecting the uncertainties of the parameters may have unexpected state estimate errors and even lead to diverge. Many methods have been proposed to solve these uncertain model parameters, such as $H_\infty$ filtering[7], desensitized Kalman filtering[8], set-valued estimation[9] and consider Kalman filtering (also called Schmidt-Kalman filtering)[6, 10, 11]. A *consider* method is proposed by Schmidt to account for the parameter uncertainties by incorporating the covariance of the parameters into the Kalman filtering formulations.

To overcome the drawbacks of the EnKF coming from the unknown parameters, this paper proposes an ensemble consider Kalman filter (EnCKF) for the nonlinear dynamic systems with uncertain parameters. The covariance matrices between the uncertain parameters and the states and the measurements are computed and propagated by using the ensemble integration in the filtering algorithm. The formula of the EnCKF are derived by using augmented-state methods[12] in Section Ⅱ. Two numerical simulations are shown in Section Ⅲ.

## II. ENSEMBLE CONSIDER KALMAN FILTERING

For the nonlinear dynamic system model, the EnKF gives a suboptimal solution of the Fokker-Planck equation by using an ensemble integration to approximate the error statistics. Here, based on the *consider* method and the EnKF, the EnCKF method is presented to consider the uncertain parameters in the dynamic models.

Consider a nonlinear discrete dynamic system model with uncertain parameters and additive noises, in which its state equation is given by

$$x_k = f(x_{k-1}, b) + w_{k-1}, \qquad (1)$$

and measurement equation is given by

$$z_k = h(x_k, b) + v_k \qquad (2)$$

where $x_k \in \mathbb{R}^n$ is the state vector at time step $k$, and $z_k \in \mathbb{R}^p$ is the measurement vector, functions $f$ and $h$ are respectively the nonlinear dynamic and the measurement equations, $b \in \mathbb{R}^l$ is referred to as the uncertain parameter vector, and $w_k, v_k$ are assumed to be zero-mean Gaussian white noise with covariance matrices $Q_k$ and $R_k$, respectively. Moreover, $w_k, v_k$ are assumed to be uncorrelated.

*Resrach supported by the National Nature Science Foundation of China under grant #61603346, the Key Science and Technology Program of Henan Province under grant #182102110014, the Fundamental Research Programs for the Provincial Universities under grant #15KYYWF01, and the Key research projects of Henan higher education institutions under grant #18A413003.

Tai-shan Lou is with School of Electrical and Information Engineering, Zhengzhou University of Light Industry, Zhengzhou, 450002 China (corresponding author: 86-370-6355-6790; e-mail: tayzan@sina.com, loutaishan@zzuli.edu.cn).

Nan-hua Chen is with School of Electrical and Information Engineering, Zhengzhou University of Light Industry, Zhengzhou, 450002 China (e-mail: chennanhua1@foxmail.com).

Hua Xiong is with Beijing Institute of Electronic System Engineering, Beijing 100854 China (e-mail: xiong_hua_@126.com).

Ya-xi Li is with School of Electrical and Information Engineering, Zhengzhou University of Light Industry, Zhengzhou, 450002 China (e-mail: liyaxi321@foxmail.com).

Lei Wang is with School of Automation Science and Electrical Engineering, Beihang University, Beijing, 100083 China. (e-mail: lwang @buaa.edu.cn).

In this work, the uncertain parameters are modeled as a constant vector with a *priori* known statistic. Here, the reference values and the covariance are assumed to be $\bar{b}$ and $Q_b$.

Following the *consider* approach, the estimated state $x_k$ and the uncertain parameters $b$ are augmented into a state $X_k \in \mathbb{R}^{n+l}$, which is described by

$$X_k = [x_k, b_k]^T \quad (3)$$

and its covariance are defined by

$$P_{XX} = \begin{bmatrix} P_{xx} & P_{xb} \\ P_{bx} & P_{bb} \end{bmatrix} \quad (4)$$

To propagate the error distribution in the predictive step, $m$ forecasted state estimates with sample random errors are generated as in an ensemble at time step $k-1$. The ensemble $\chi_{k-1}^f \in R^{n \times m}$ is defined as

$$\chi_{k-1}^f = \{X_{k-1}^1, X_{k-1}^2, \cdots, X_{k-1}^m\} \quad (5)$$

where the superscript $i$ denote the $i$-th forecast ensemble member, the augmented ensemble member is $X_{k-1}^i = [x_{k-1}^i, b_{k-1}^i]^T$, $i = 1, 2, \cdots, m$.

The propagated ensemble members are

$$X_k^i = \begin{bmatrix} f(X_{k-1}^i) \\ \hat{b}_{k-1}^i \end{bmatrix} + \begin{bmatrix} w_k^i \\ 0 \end{bmatrix} \quad (6)$$

Estimating the priori state $\hat{X}_k^-$ by

$$\hat{X}_k^- = \begin{bmatrix} \hat{x}_k^- \\ \hat{b}_k^- \end{bmatrix} = \frac{1}{m} \sum_{i=1}^{m} X_k^i \quad (7)$$

and the ensemble error matrix $M_k^X \in \mathbb{R}^{(n+l) \times m}$, which is the difference between the true state $X_k^i (i=1,2,\cdots,m)$ and the ensemble mean $\hat{X}_k^-$, is defined by

$$M_k^X = \begin{bmatrix} M_k^x \\ M_k^b \end{bmatrix} = [X_k^1 - \hat{X}_k^-, \cdots, X_k^m - \hat{X}_k^-] \quad (8)$$

From Eqs. (7) and (8), it can be seen that

$$\hat{x}_k^- = \frac{1}{m} \sum_{i=1}^m x_k^i \quad (9)$$

$$\hat{b}_k^- = \frac{1}{m} \sum_{i=1}^m b_k^i \quad (10)$$

$$M_k^x = [x_k^1 - \hat{x}_k^-, \cdots, x_k^m - \hat{x}_k^-] \quad (11)$$

$$M_k^b = [b_k^1 - \hat{b}_k^-, \cdots, b_k^m - \hat{b}_k^-] \quad (12)$$

where $\hat{x}_k^- \in \mathbb{R}^n$ is the ensemble mean of $m$ state $x_k$, $\hat{b}_k^- \in \mathbb{R}^l$ is the ensemble mean of $m$ parameter $b_k$. The ensemble error matrix $M_k^x \in \mathbb{R}^{n \times m}$ is the difference between the ensemble members and the ensemble mean, and the ensemble $M_k^b \in \mathbb{R}^{q \times m}$ is the difference between ensemble error matrix of $b_k^i$ and $\hat{b}_k^-$, which is replaced by the constant $\bar{b}$ of the uncertain parameter in *consider* approach to reduce the computation.

Estimating the predicted covariance matrix $P_{XX,k}^-$, which is given by the estimated and consider components, and is defined as

$$\begin{aligned} P_{XX,k}^- &= \begin{bmatrix} P_{xx,k}^- & P_{xb,k}^- \\ P_{bx,k}^- & P_{bb,k}^- \end{bmatrix} \\ &= \frac{1}{m-1} M_k^X (M_k^X)^T \\ &= \frac{1}{m-1} \begin{bmatrix} M_k^x \\ M_k^b \end{bmatrix} [(M_k^x)^T \quad (M_k^b)^T] \\ &= \frac{1}{m-1} \begin{bmatrix} M_k^x (M_k^x)^T & M_k^x (M_k^b)^T \\ M_k^b (M_k^x)^T & M_k^b (M_k^b)^T \end{bmatrix} \end{aligned} \quad (13)$$

from which it can be seen that

$$\begin{aligned} P_{xx,k}^- &= \frac{1}{m-1} M_k^x (M_k^x)^T \\ P_{xb,k}^- &= \frac{1}{m-1} M_k^x (M_k^b)^T \\ P_{bb,k}^- &= \frac{1}{m-1} M_k^b (M_k^b)^T \end{aligned} \quad (14)$$

The measurement ensemble members $Z_k^i (i=1,2,\cdots,m)$ are computed from the augmented state ensemble members $X_k^i (i=1,2,\cdots,m)$:

$$Z_k^i = h(X_k^i) = h(x_k^i, b_k^i) \quad (15)$$

Estimating the priori measurement $\hat{z}_k^-$ by

$$\hat{z}_k^- = \frac{1}{m} \sum_{i=1}^m Z_k^i \quad (16)$$

and the measurement error matrix $M_k^z \in R^{p \times m}$ is defined by

$$M_k^z = [Z_k^1 - \hat{z}_k^-, \cdots, Z_k^m - \hat{z}_k^-] \quad (17)$$

The innovations covariance $P_{zz,k}$ and the cross-covariance $P_{Xz,k}$ of the augmented measurement and state are computed by using the state and measurement ensemble members

$$P_{zz,k} = \frac{1}{m-1} M_k^z (M_k^z)^T \quad (18)$$

$$P_{Xz,k} = \frac{1}{m-1} M_k^X (M_k^z)^T \quad (19)$$

The measurement and state cross covariance $P_{Xz,k}$ in Eq. (19) is defined by the estimated and considered terms

$$P_{Xz,k} = \begin{bmatrix} P_{xz,k} \\ P_{bz,k} \end{bmatrix} = \frac{1}{m-1} \begin{bmatrix} M_k^x \\ M_k^b \end{bmatrix} (M_k^z)^T \quad (20)$$

it can be seen that

$$P_{xz,k} = \frac{1}{m-1} M_k^x (M_k^z)^T \quad (21)$$

$$P_{bz,k} = \frac{1}{m-1} M_k^b (M_k^z)^T \quad (22)$$

Then, the augmented Kalman gain is defined as:

$$K_k = \begin{pmatrix} K_{x,k} \\ K_{b,k} \end{pmatrix} = P_{Xz,k} P_{zz,k}^{-1} = \begin{bmatrix} P_{xz,k} \\ P_{bz,k} \end{bmatrix} P_{zz}^{-1} \quad (23)$$

it can be seen that

$$K_{x,k} = P_{xz,k} P_{zz,k}^{-1} \quad (24)$$

$$K_{b,k} = P_{bz,k} P_{zz,k}^{-1} \quad (25)$$

Then, the augmented posteriori estimate is

$$\hat{X}_k^{+i} = \hat{X}_k^{-i} + K_k (z_k^i - Z_k^i) \quad (26)$$

and the augmented posteriori covariance matrix is

$$P_{XX,k}^+ = \begin{bmatrix} P_{xx,k}^+ & P_{xb,k}^+ \\ P_{bx,k}^+ & P_{bb,k}^+ \end{bmatrix}$$
$$= \begin{bmatrix} P_{xx,k}^- & P_{xb,k}^- \\ P_{bx,k}^- & P_{bb,k}^- \end{bmatrix} - \begin{bmatrix} K_{x,k} P_{zz,k} K_{x,k}^T & K_{x,k} P_{zz,k} K_{b,k}^T \\ K_{b,k} P_{zz,k} K_{x,k}^T & K_{b,k} P_{zz,k} K_{b,k}^T \end{bmatrix} \quad (27)$$

where the perturbed measurements $z_k^i$ are given by

$$z_k^i = h(x_k, b) + v_k^i \quad (28)$$

in which the perturbation variable $v_k^i \sim N(0, R_k)$.

It can be seen that the above formulations is a general form of the EnKF algorithm. In *consider* approach, the gain matrix $K_{b,k}$ of the uncertain parameter is forced to be zero, that is to say, $K_{b,k} = 0$. Following this consider approach, the posteriori estimate and covariance matrix are respectively

$$\hat{X}_k^{+i} = \hat{X}_k^{-i} + \begin{pmatrix} K_{x,k} \\ 0 \end{pmatrix} (z_k^i - Z_k^i) \quad (29)$$

$$P_{XX,k}^+ = \begin{bmatrix} P_{xx,k}^- & P_{xb,k}^- \\ P_{bx,k}^- & P_{bb,k}^- \end{bmatrix} - \begin{bmatrix} K_{x,k} P_{zz,k} K_{x,k}^T & K_{x,k} P_{bz,k}^T \\ P_{bz,k} K_{x,k}^T & 0 \end{bmatrix} \quad (30)$$

from which it can be seen that

$$\hat{x}_k^{+i} = \hat{x}_k^{-i} + K_{x,k} (z_k^i - Z_k^i) \quad (31)$$

$$P_{xx,k}^+ = P_{xx,k}^- - K_{x,k} P_{zz,k} K_{x,k}^T \quad (32)$$

$$P_{xb,k}^+ = P_{xb,k}^- - K_{x,k} P_{bz,k}^T \quad (33)$$

$$P_{bb,k}^+ = P_{bb,k}^- = Q_b \quad (34)$$

where the last formulation of Eq. (30) is obtained by the substituting Eq.(25) into the second formulation of the Eq.(27).

Lastly, the posteriori estimate and the posteriori parameter can be given by

$$\hat{x}_k^+ = \frac{1}{m} \sum_{i=1}^m \hat{x}_k^{+i} \quad (35)$$

$$\hat{b}_k^+ = \hat{b}_k^- \quad (36)$$

Note that the standard EnKF gain matrix can be recovered by setting $Q_b = 0$.

Following above formulations, the ensemble consider Kalman filter is summarized by the following equations, which include two parts:

Time update:

$$\begin{aligned}
x_k^i &= f(\hat{x}_{k-1}^i, b_{k-1}^i) \\
\hat{x}_k^- &= \frac{1}{m} \sum_{i=1}^m x_k^i \\
P_{xx,k}^- &= \frac{1}{m-1} M_k^x (M_k^x)^T \\
P_{xb,k}^- &= \frac{1}{m-1} M_k^x (M_k^b)^T
\end{aligned} \quad (37)$$

Measurement update:

$$\begin{aligned}
\hat{z}_k^- &= \frac{1}{m} \sum_{i=1}^m Z_k^i \\
P_{zz,k} &= \frac{1}{m-1} M_k^z (M_k^z)^T \\
P_{xz,k} &= \frac{1}{m-1} M_k^x (M_k^z)^T \\
P_{bz,k} &= \frac{1}{m-1} M_k^b (M_k^z)^T
\end{aligned} \quad (38)$$

$$\begin{aligned}
K_{x,k} &= P_{xz,k} P_{zz,k}^{-1} \\
\hat{x}_k^{+i} &= \hat{x}_k^{-i} + K_{x,k} (z_k^i - Z_k^i)
\end{aligned} \quad (39)$$

$$\hat{x}_k^+ = \frac{1}{m} \sum_{i=1}^m \hat{x}_k^{+i}$$

$$\begin{aligned}
P_{xx,k}^+ &= P_{xx,k}^- - K_{x,k} P_{zz,k} K_{x,k}^T \\
P_{xb,k}^+ &= P_{xb,k}^- - K_{x,k} P_{bz,k}^T
\end{aligned} \quad (40)$$

where

$$\begin{aligned}
M_k^x &= [x_k^1 - \hat{x}_k^-, \cdots, x_k^m - \hat{x}_k^-] \\
M_k^z &= [z_k^1 - \hat{z}_k^-, \cdots, z_k^m - \hat{z}_k^-] \\
M_k^b &= [b_k^1 - b_0, \cdots, b_k^m - b_0]
\end{aligned} \quad (41)$$

$$\begin{aligned}
Z_k^i &= h(x_k^i, b_k^i) \\
z_k^i &= h(x_k, b) + v_k^i
\end{aligned} \quad (42)$$

**Note**: From the last time step $k-1$ to the next time step $k$, the standard EnKF does not change the ensemble members, but in this consider approach, then the estimates of the uncertain parameters are discarded, and the cross-covariance between the uncertain parameters and the states must be considered into the state estimate error matrices. However, the standard EnKF algorithm does not propagate the covariance, so the above cross-covariance can't be reflected into the EnKF algorithm.

For considering the uncertain parameters in the dynamic system model, the ensemble members at time step $k-1$ should be resampled by introduced the augmented covariance matrix $P_{XX,k-1}$. Firstly, the augmented covariance matrix $P_{XX,k-1}$ should be factorized by the Cholesky factorization to obtain its lower-diagonal square-root factorization $S_{XX,k-1}$. That is to say, $P_{XX,k-1} = S_{XX,k-1} S_{XX,k-1}^T$, in which $S_{XX,k-1}$ can be evaluated by the following formulation

$$S_{XX,k-1} = \begin{pmatrix} S_{xx,k-1} & 0 \\ P_{xb,k-1}^T (S_{xx,k-1}^-)^T & \sqrt{Q_b - P_{xb,k-1}^T (P_{xx,k-1}^+)^{-1} P_{xb,k-1}} \end{pmatrix} \quad (43)$$

where $P_{xx,k-1} = S_{xx,k-1} S_{xx,k-1}^T$. Then, the new ensemble members can be generated by using the square-root matrix $S_{XX,k-1}$ such that $X_{k-1}^i \sim N(\hat{X}_{k-1}^+, S_{XX,k-1}) (i=1,2,\cdots,m)$.

### III. SIMULATION AND ANALYSIS

To evaluate the effectiveness of the presented EnCKF algorithm in the section II. Two numerical simulations with uncertain parameters are considered to compare the EnCKF with the classical EnKF.

#### A. Spacecraft attitude tracking system

The spacecraft attitude tracking system is used to track the spacecraft drift signal, but it always has uncertain parameters, or unknown biases. The measurement information is provided by the gyroscope sensor. The discrete dynamic systems with the states $x = [x1, x2]^T$ are modeled by[11]

$$x_{k+1} = \begin{bmatrix} 0 & 1 \\ -0.85 & 1.70 \end{bmatrix} x_k + \begin{bmatrix} 0.0129 \\ -1.2504 \end{bmatrix} b_k + \begin{bmatrix} 0 \\ 1 \end{bmatrix} w_k \quad (44)$$

$$z_k = \begin{bmatrix} 0 & 1 \end{bmatrix} x_k + v_k \quad (45)$$

The true state is $x_0 = [2 \quad 1]^T$, the uncertain parameter is set as $b_0 \sim N(0, 0.5^2)$, the dynamic noises and measurement noises are the Gaussian distribution with zero mean, and their covariance are $0.05^2$ and $0.5^2$, respectively.

In simulations, the initial values of the state estimation and the covariance are respectively $\hat{x}_0 = [2 \quad 1]^T$ and $P_0 = 0.025$. The reference value of the uncertain parameter is set as $\hat{b}_0 = 0$ in the classical EnKF algorithm. Each single simulation time-step is 40s. One hundred Monte Carlo runs, and the root mean squared errors (RMSE) of the state estimate at each epoch are computed to evaluate the performance of the EnCKF with the EnKF.

The two state estimate RMSEs of the EnCKF and the EnKF, in which the number of the ensemble member of the two filters are all 13 and 21, are shown in Figs. 1 and 2. It can be seen that the RMSEs of the EnCKF are much smaller than the RMSEs of the EnKF as a whole. When the ensemble members increase from 13 to 21, the two RMSEs of the estimate errors decrease. In a word, when the uncertain parameters, or the unknown biases, of the dynamic system cannot be accurately modeled, the classical EnKF has a bad performance in simulation, and the proposed EnCKF algorithm can mitigate the negative effects of the uncertain parameters.

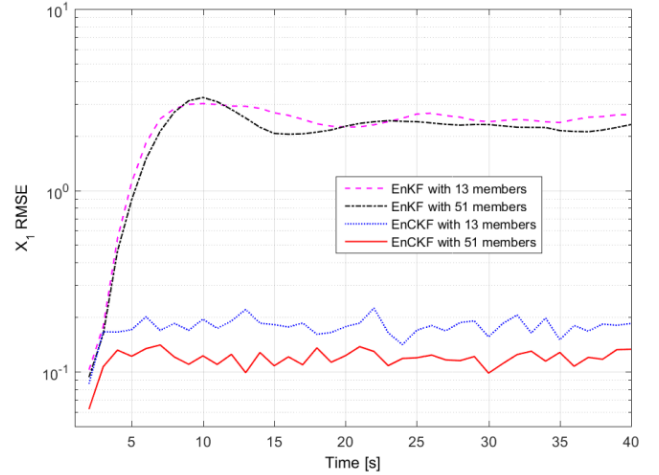

Figure 1. RMSEs of the first state

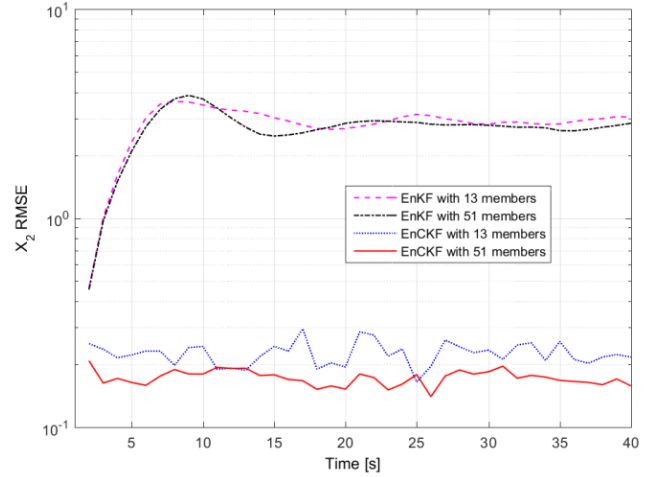

Figure 2. RMSEs of the second state

#### B. Univariate non-stationary growth model

The univariate non-stationary growth model is highly nonlinear example, and is always used to evaluate performance of the nonlinear filtering algorithm[1]. It's dynamic state space model is defined by

$$x_k = 0.5 x_{k-1} + \frac{2.5 x_{k-1}}{1 + x_{k-1}^2} + 8\cos(1.2(k-1)) + w_{k-1} \quad (46)$$

and measurements equation

$$z_k = \frac{x_k^2}{20} + b_k + v_k \quad (47)$$

where $w_{k-1}$ and $v_k$ are independent zero-mean Gaussian white noise with covariance matrices $Q_{k-1}=1$ and $R_k=1$, respectively. The initial state and covariance are given as $x_0=0$ and $P_0=10$, and the uncertain parameter (or called unknown bias) $b$ follows a normal distribution with mean $b_0=5$ and covariance $Q_b=10^2$.

In simulations, the total number of simulation is set at 200 steps; each single run lasts for 50 Monte Carlo for the the EnCKF algorithm and the EnKF algorithm, in which the ensemble members of the state are set as 13 and 51 in the two filters.

Fig. 3 shows the simulation results, which are the RMSEs for two filters. In Fig. 3, the mean of the RMSEs of the EnKF and the EnCKF are respectively1.8222 and 1.3904, when the ensemble members are 13; the mean of the RMSEs of the two filters are respectively 1.7768 and 1.2443, when the ensemble members are 51.

From Fig. 3 and the means of the RMSEs, it can be seen that the RMSEs for the EnCKF are all smaller than that of the EnKF, and the RMSEs of the ensemble members of 51 is smaller than that of 13 when using the EnCKF algorithm in the simulation. In other word, the proposed EnCKF can reduce the negative effects of the unknown biases compared with the EnKF when the information of the unknown biases is incomplete, and it can improve the accuracy of EnCKF for the highly nonlinear systems by increasing number of ensemble members.

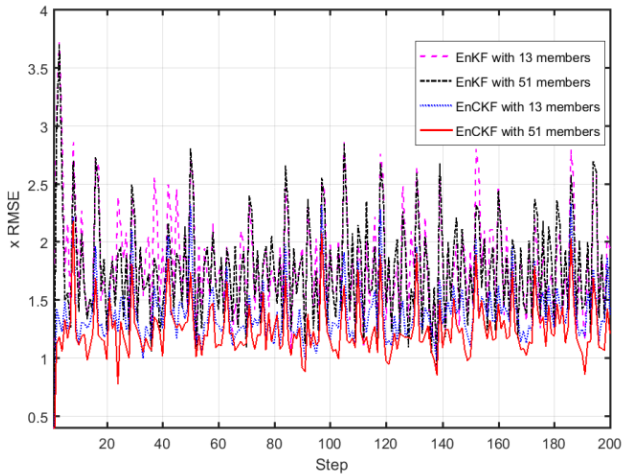

Figure 3.   RMSEs of the state

## IV. CONCLUSION

In this paper, the ensemble consider Kalman filter is proposed to mitigate the negative effects of uncertain parameters in nonlinear dynamic and measurement models. The ensemble Kalman filter can avoid using the Jacobian matrices and reduce the computational complexity, the unknown parameters of the models still are not considered. By incorporating the statistics of the uncertain parameters into the state estimation formulations and using an augmented-state approach, the ensemble integration is reset by resampling the ensemble members in the new step, and the EnCKF algorithm is derived. Two numerical simulations show that the presented EnCKF can mitigate the negative effects of the uncertain parameters.